\documentclass[twocolumn,prd,unsortedaddress,superscriptaddress,showpacs,a4paper,nofootinbib]{revtex4-1} 
\usepackage{epsfig} 
\usepackage{subfigure} 
\usepackage{amsmath} 
\usepackage[dvips]{color} 
 
\def\beq{\begin{equation}} 
\def\enq{\end{equation}} 
\def\beqa{\begin{eqnarray}} 
\def\enqa{\end{eqnarray}}

\def\qq{\lag\bar{q}q\rag} 
\def\uu{\lag\bar{u}u\rag} 
\def\dd{\lag\bar{d}d\rag}

\def\G3{\lag g^3G^3\rag}

\def\al{\alpha}

\def\lb{\label} 
\def\nnb{\nonumber} 
\def\nn{\nonumber}

\newcommand{\rag}{\rangle} 
\newcommand{\lag}{\langle}

\def\MeV{\mbox{ MeV}} 
\def\GeV{\mbox{ GeV}} 
\begin{document} 
 
\title{\sc QCD Sum Rules for the  production of the X(3872)  as a 
  mixed molecule-charmonium state in B meson decay} 
 
\author{C.M.~Zanetti}   \email{carina@if.usp.br}   \author{M.~Nielsen} 
\email{mnielsen@if.usp.br}   \affiliation{Instituto   de  F\'{\i}sica, 
  Universidade de S\~{a}o Paulo, C.P.  66318, 05389-970 S\~{a}o Paulo, 
  SP,    Brazil}   \author{R.~D.~Matheus}   \email{matheus@unifesp.br} 
\affiliation{Departamento   de   Ci\^encias   Exatas   e   da   Terra, 
  Universidade Federal de S\~ao Paulo, 09972-270, Diadema, SP, Brazil} 
 
\begin{abstract} 
 
  We  use QCD  sum  rules to  calculate  the branching  ratio for  the 
  production  of the  meson $X(3872)$  in the  decay  $B\to X(3872)K$, 
  assumed  to be  a mixture  between charmonium  and  exotic molecular 
  $[c\bar{q}][q\bar{c}]$ states with $J^{PC}=1^{++}$.  We find that in 
  a   small   range   for    the   values   of   the   mixing   angle, 
  $5^\circ\leq\theta\leq13^\circ$,     we    get    the     branching    ratio 
  ${\mathcal{B}}(B\to  XK)=(1.00\pm0.68)\times10^{-5}$,  which  is  in 
  agreement  with  the  experimental  upper  limit.   This  result  is 
  compatible with the analysis of the mass and decay width of the mode 
  $J/\psi(n\pi)$ and the radiative decay mode $J/\psi\gamma$ performed 
  in the same approach. 
 
\end{abstract}  
 
\maketitle 
 
%%%%%%%%%%%%%%%%%%%%%%%%%%%%%%%%%%%%%%%%%%%%%%%%%%%%%%%%%%%%%%%%%%%%% 
% 
% 
\section{Introduction} 
%  
%  
%%%%%%%%%%%%%%%%%%%%%%%%%%%%%%%%%%%%%%%%%%%%%%%%%%%%%%%%%%%%%%%%%%%%% 
 
The $X(3872)$ state has been first observed by the Belle Collaboration 
in                              the                              decay 
$B^+\!\rightarrow\!X(3872)K^+\rightarrow\!J/\psi\pi^+\pi^-         K^+$ 
\cite{belle1},  and   was  later  confirmed  by  CDF,   D0  and  BaBar 
\cite{Xexpts}.  This  was the first  state of an increasing  number of 
candidates for exotic hadrons  discovered recently.  The current world 
average mass is $m_X=(3871.4\pm0.6)\MeV$, and the width is $\Gamma<2.3 
\MeV$  at 90\%  confidence  level.  BaBar  Collaboration reported  the 
radiative decay mode  $X(3872)\to \gamma J/\psi$ \cite{belleE,babar2}, 
which  determines $C=+$.   Further  studies from  Belle  and CDF  that 
combine   angular  information   and  kinematic   properties   of  the 
$\pi^+\pi^-$ pair, strongly favors the quantum numbers $J^{PC}=1^{++}$ 
or $2^{-+}$ \cite{belleE,cdf2,cdf3}.  Although the new BaBar result 
favors the $J^{PC}=2^{-+}$ assignment \cite{newbabar}, established properties 
of the $X(3872)$ are in conflict with this assignment \cite{kane,bpps}. 
Therefore, in this work we will consider the $X(3872)$ as being a  
$J^{PC}=1^{++}$ state. BaBar Collaboration reported the 
upper limit  of the  branching ratio for  the production in  $B$ meson 
decay \cite{Aubert2005vi}: 
\begin{equation} 
  \label{branching} 
  {\mathcal B}(B^\pm\to K^\pm X(3872))<3.2\times10^{-4}. 
\end{equation} 
Recently, Belle Collaboration presented the most precise measurement of 
the         branching         fraction          ${\mathcal{B}}(B^\pm\to 
X(3872)K^\pm){\mathcal{B}}(X(3872)\gamma 
J/\psi)=(1.78^{+0.48}_{-0.44}\pm0.12)\times10^{-6}$ \cite{Bellenew}. 
 
The decay  modes of the  $X(3872)$ into $J/\psi$ and  other charmonium 
states indicate the existence of a $\bar{c}c$ in its content.  However 
the attempts  to classify the  state in the   charmonium spectrum 
have  to deal with  the fact  that the  mass of  the $X(3872)$  is not 
compatible  with any  of the  possible candidates  in the  quark model 
\cite{bg}.  Another  problem comes from  the measurement of  the decay 
rates  of  the  processes  $X(3872) \to  J/\psi\,\pi^+\pi^-\pi^0$  and 
$X(3872)\rightarrow\!J/\psi\pi^+\pi^-$,     which    are    comparable 
\cite{belleE}, 
\beq    {X       \to       J/\psi\,\pi^+\pi^-\pi^0\over 
    X\to\!J/\psi\pi^+\pi^-}=1.0\pm0.4\pm0.3\,, 
\label{ratio} 
\enq 
that could indicate a strong  isospin and G parity violation,  which  is 
incompatible  with a  $c\bar{c}$  structure.  
 
The  coincidence  between  the  $X(3872)$  mass  and  the  $D^{*0}D^0$ 
threshold:  $M(D^{*0}D^0)=(3871.81\pm0.36)\MeV$  \cite{cleo}, inspired 
the   proposal    that   the   $X(3872)$   could    be   a   molecular 
$(D^{*0}\bar{D}^0-\bar{D}^{*0}D^0)$  bound  state  with small  binding 
energy \cite{close,swanson}. In particular, considering the $X(3872)$   
as an  admixture of neutral and charged components of molecules, i.e.   
$D^{0}\bar{D}^{*0}$ and $D^+D^{*-}$, the strong isospin violation  
observed in Eq.~(\ref{ratio}) 
could be explained in a  very natural way \cite{oset2,nnl}.  
There is also  a possibility that the observed ratio of 
the  $X$ decaying into  $J/\psi+2\pi$ or  $3\pi$ may  not come  from a 
large isospin breaking.  In  Ref.~\cite{oset1} the isospin breaking is 
investigated in  the dynamical  generation of the  $X$ as  a molecular 
state, and it is found to be small. But, considering that the two pion 
and three  pion states  comes from the  decays of $\rho$  and $\omega$ 
mesons, the small  isospin breaking is compensated by  the larger phase 
space  of the  $\rho$ meson,  thus explaining  the  experimental data. 
 
  Another interesting interpretation for the $X(3872)$ 
is    that    it    could    be    a    compact    tetraquark    state 
\cite{maiani,tera,tera2}. 
 
The  co-existence  of both  $\bar{c}c$  and  multiquark components  is 
subject  of  debate  in  many  works,  and it  is  supported  by  some 
experimental data.  In  Ref.~\cite{Bignamini:2009sk}, a simulation for 
the production  of a bound $D^0\bar{D}^{*0}$ state  with biding energy 
as small as  0.25 MeV, reported a production cross  section that is an 
order of  magnitude smaller than  the cross section obtained  from the 
CDF data.  A similar  result was obtained  in Ref.~\cite{suzuki}  in a 
more   phenomenological   analysis.  However,   as   pointed  out   in 
Ref.~\cite{Artoisenet:2009wk},   a    consistent   analysis   of   the 
$D^0\bar{D}^{*0}$ molecule production requires taking into account the 
effect of final state interactions of the $D$ and $D^*$ mesons. 
  
Besides, the recent observation,  reported by BaBar \cite{babar09}, of 
the decay  $X(3872)\to \psi(2S)\gamma$  at a rate:  $ {{\cal  B}(X \to 
  \psi(2S)\,\gamma)\over {\cal  B}(X\to\psi\gamma)}=3.4\pm1.4, $ it is 
much bigger  than the molecular prediction  ~\cite{swan1}: $ {\Gamma(X 
  \to          \psi(2S)\,\gamma)\over\Gamma(X\to\psi\gamma)}\sim4\times 
10^{-3}.$  
 
In the framework of the QCD sum rules (QCDSR) the mass of the $X$ were 
computed  with  good   agreement  with  data,  considering  tetraquark 
\cite{x3872} and molecular structures  \cite{nnl}. The same success is 
not  found in  decay widths  calculations.  In  Ref.~\cite{decayx} the 
decay  width  of  the  modes  $J/\psi(n\pi)$ are  calculated  for  the 
tetraquark  structure, and  the result  is one  order larger  that the 
total width.  In  Ref.~\cite{x3872mix,x3872rad} the QCDSR approach was 
used to study the $X(3872)$ structure including the possibility of the 
mixing between  two and four-quark  states, where it  was successfully 
applied to obtain the mass of  the state and the decays widths for the 
modes  $J/\psi(n\pi)$  and the  radiative  decay mode  $J/\psi\gamma$. 
This   was  implemented  following   the  prescription   suggested  in 
\cite{oka24} for the light sector.  The mixing is done at the level of 
the  currents and is  extended to  the charm  sector.  In  a different 
context (not  in QCDSR), a  similar mixing was suggested  already some 
time ago  by Suzuki \cite{suzuki}.  Physically, this  corresponds to a 
fluctuation of the $c \overline{c}$ state where a gluon is emitted and 
subsequently splits into a light quark-antiquark pair, which lives for 
some time and behaves like a molecule-like state 
  
The models  for the quark structure  can also be applied  to study the 
production of  the state  in $B$ decays.   This subject is  studied in 
different approaches in  Refs.~\cite{bxk1,bxk2,bxk3}.  In this work we 
will focus on  production of the $X(3872)$, using  the mixed two-quark 
and   four-quark  prescription  of   Ref.~\cite{x3872mix,x3872rad}  to 
perform   a  QCDSR  analysis   of  the   process  $B^\pm\to 
X(3872)K^\pm$. 
 
\section{The decay  $B\to X(3872)K$} 
 
\begin{figure}[h]  
\centerline{ 
\epsfig{width=0.3\textwidth,figure=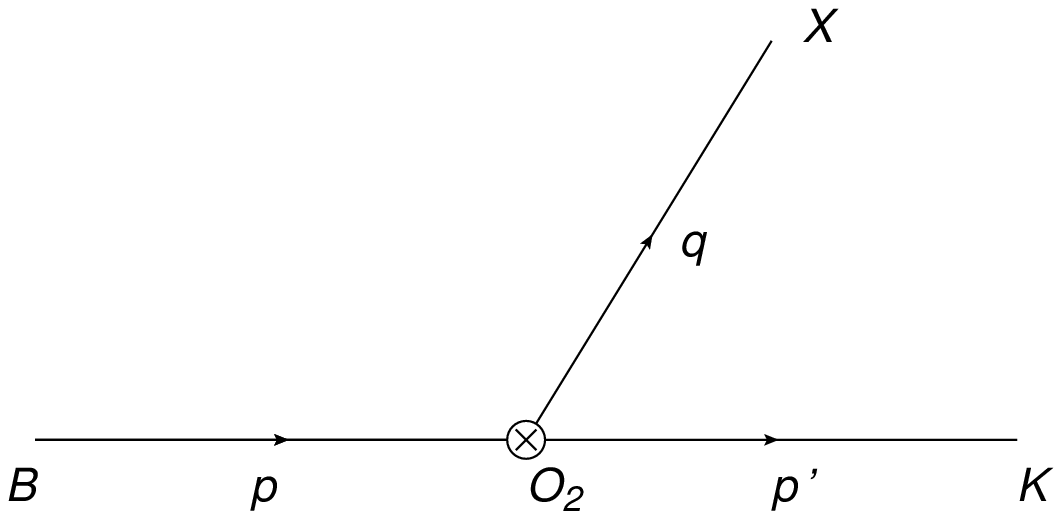} 
} 
\caption{\label{fig1}} 
\end{figure} 
 
The process  $B\to X(3872)K$ occurs via  weak decay of  the $b$ quark, 
while the $u$ quark  is a spectator. The $X$ meson 
as a mixed  state of molecule and charmonium  interacts via $\bar{c}c$ 
component  of the  weak current.   In effective  theory, at  the scale 
$\mu\sim m_b\ll m_W$, the weak  decay is treated as a four-quark local 
interaction described by the effective Hamiltonian (see Fig.~\ref{fig1}): 
\begin{equation}\lb{ham} 
  {\mathcal{H}}_W=\frac{G_F}{\sqrt{2}}V_{cb}V_{cs}^*\left[\left(C_2(\mu)+ 
\frac{C_1(\mu)}{3}\right) 
    {\mathcal{O}}_2+\cdots\right]\,, 
\end{equation} 
where $V_{ik}$ are CKM  matrix elements, $C_1(\mu)$ and $C_2(\mu)$ are 
short  distance  Wilson coefficients  computed  at the  renormalization 
scale $\mu\sim{\mathcal O}(m_b)$.   The four-quarks effective operator 
is ${\mathcal{O}}_2=(\bar{c}\Gamma_\mu  c)(\bar{s}\Gamma^\mu b)$, with 
$\Gamma_\mu=\gamma_\mu(1-\gamma_5)$. 
 
The decay amplitude of the process is calculated from the Hamiltonian  
(\ref{ham}), 
and it can be factorized by splitting the matrix element in two pieces: 
\begin{eqnarray}\lb{amp} 
{\mathcal M} 
&=&i\frac{G_F}{\sqrt{2}}V_{cb}V_{cs}^*\left(C_2+\frac{C_1}{3}\right)\nn\\& 
\times&\langle B(p)\vert J_{\mu}^W\vert K(p^\prime)\rangle\langle X(q) 
\vert J^{\mu(\bar{c}c)}\vert0\rangle, 
\end{eqnarray} 
where      $p=p^\prime+q$,      and      the     currents  are  
\begin{equation}\lb{wcurrents}    J_{\mu}^W=\bar{s}\Gamma_\mu   b\,,\quad 
  J_{\mu}^{(\bar{c}c)}=\bar{c}\Gamma_\mu  c\,.\end{equation}  The  matrix 
elements in Eq.~({\ref{amp}}) are parametrized in the following way 
\begin{equation}\lb{2pmatrix} 
\langle X(q)\vert J_{\mu}^{(\bar{c}c)}\vert0\rangle=\lambda_W 
\epsilon^*_\mu(q)\,, 
\end{equation} 
and 
\begin{equation}\lb{3pmatrix} 
\langle B(p)\vert J_{\mu}^W\vert K(p^\prime)\rangle=f_+(q^2)(p_\mu+ 
p_\mu^\prime)+f_-(q^2)(p_\mu-p_\mu^\prime)\,. 
\end{equation} 
The  parameter  $\lambda_W$  in  (\ref{2pmatrix}) gives  the  coupling 
between the current $J_\mu^{(\bar{c}c)}$  and the $X$ state.  The form 
factors $f_\pm(q^2)$  describe the weak transition $B\to  K$. Hence we 
can see  that the  factorization of the  matrix element  describes the 
decay as two separated sub-processes. 
 
The decay width for the process $B^\pm\to X(3872)K^\pm$ is given by  
\begin{equation}\lb{eqwidth} 
\Gamma(B\to XK)=\frac{1}{16\pi m_B^3}\lambda^{1/2}(m_B^2,m_K^2,m_X^2) 
\vert{\mathcal{M}}\vert^2, 
\end{equation} 
with $\lambda(x,y,z)=x^2+y^2+z^2-2xy-2xz-2yz$. The invariant amplitude 
squared can be obtained from (\ref{amp}): 
 
\begin{eqnarray} 
\vert\mathcal{M}\vert^2&=&\frac{G_F^2}{2}\vert V_{cb}V_{cs}\vert^2\left(C_2 
+\frac{C_1}{3}\right)^2\nn\\\nn\\ 
&\times& \lambda(m_B^2,m_K^2,m_X^2)\lambda_W^2f_+^2(q^2)\vert_{q^2\to-m_X^2} 
\,. 
\end{eqnarray} 
We  will  use  QCD sum  rules  in  order  to determine  the  parameter 
$\lambda_W$  and the  form  factor $f_+(q^2)$,  and  therefore we  can 
obtain the width of the decay.

\section{Two-point correlator} 
 
The QCD sum rule approach \cite{svz,rry,SNB} is based on the two point  
correlator: 
\begin{equation} 
\Pi_{\mu\nu}(q)=i\int d^4y~e^{iq\cdot y}\langle0\vert T\{J_\mu^X(y) 
J_\nu^{(\bar{c}c)}\}\vert0\rangle\,, 
\end{equation} 
where    the    current    $J_\nu^{(\bar{c}c)}$    is    defined    in 
(\ref{wcurrents}).  For the $X$  meson we will follow \cite{oka24} and 
consider    a    mixed     charmonium-molecular    current    as    in 
Ref.~\cite{x3872mix,x3872rad}.   For the  charmonium part  we  use the 
conventional axial current: 
\begin{equation} 
j'^{(2)}_{\mu}(x) = \bar{c}_a(x) \gamma_{\mu} \gamma_5 c_a(x). 
\lb{curr2} 
\end{equation} 
The  $D^0$   $D^{*0}$ molecule    is   interpolated  by 
\cite{liuliu,dong,stancu}   (with   $q=u$): 
\begin{eqnarray} 
j^{(4q)}_{\mu}(x) & = & {1 \over \sqrt{2}} 
\bigg[ 
\left(\bar{q}_a(x) \gamma_{5} c_a(x) 
\bar{c}_b(x) \gamma_{\mu}  q_b(x)\right) \nonumber \\ 
& - & 
\left(\bar{q}_a(x) \gamma_{\mu} c_a(x) 
\bar{c}_b(x) \gamma_{5}  q_b(x)\right) 
\bigg], 
\lb{curr4} 
\end{eqnarray} 
As in Ref.~\cite{oka24} we define the normalized two-quark current as 
\begin{equation} 
j^{(2q)}_{\mu} = {1 \over 6 \sqrt{2}} \uu j'^{(2)}_{\mu}, 
\lb{curr2} 
\end{equation} 
and   from  these   two  currents   we  build   the   following  mixed 
charmonium-molecular current for the $X(3872)$: 
\begin{equation} 
J_{\mu}^q(x)= \sin(\theta) j^{(4q)}_{\mu}(x) + \cos(\theta)  
j^{(2q)}_{\mu}(x).  
\lb{field} 
\end{equation} 
We  will consider  a small  admixture of  $D^+D^{*-}$  and $D^-D^{*+}$ 
components, then we  have for the $X$ current: 
\begin{equation} 
J_{\mu}^X(x)= \cos\alpha J_{\mu}^u(x)+\sin\alpha J_{\mu}^d(x), 
\label{4mix} 
\end{equation} 
with $J_{\mu}^u(x)$ and $J_{\mu}^d(x)$ given by Eq.(\ref{field}). 
 
Considering the $u$ and $d$ quarks to be degenerate, {\it i.e.}, 
$m_u=m_d$ and $\uu=\dd$, and inserting the currents (\ref{wcurrents}) 
and (\ref{4mix}) in the correlator we have in the OPE side of the sum 
rule 
\begin{eqnarray} 
\Pi^{\mathrm{OPE}}_{\mu\nu}(q)&=&(\cos\alpha+\sin\alpha)\biggl(\sin\theta\, 
\Pi^{4,2}_{\mu\nu}(q)\nn\\&+&\frac{\qq}{6\sqrt{2}}\cos\theta\, 
\Pi^{2,2}_{\mu\nu}(q)\biggr)\,, 
\end{eqnarray} 
where 
\begin{eqnarray} 
\Pi^{2,4}_{\mu\nu}(q)&=&i\int d^4y ~e^{iq\cdot y}\langle0\vert T\{ 
J_\mu^{4q}(y)J_{\nu(\bar{c}c)}(0)\}\vert0\rangle\nn\\ 
\Pi^{2,2}_{\mu\nu}(q)&=&i\int d^4y ~e^{iq\cdot y}\langle0\vert T\{ 
J_\mu^{2q}(y)J_{\nu(\bar{c}c)}(0)\}\vert0\rangle\,. 
\end{eqnarray} 
The   contribution    from   the   vector   part    of   the   current 
$J_\nu^{(\bar{c}c)}$  vanishes  after  the integration  is  performed, 
hence these  correlators are  equal (except for  a minus sign)  to the 
ones calculated  in Ref.~\cite{x3872mix} for  the two-point correlator 
of the $X(3872)$. 
 
On the  phenomenological side  the correlator is  determined inserting 
the intermediate state of the $X$: 
\begin{eqnarray} 
  \Pi_{\mu\nu}^{phen}(q)&=&\frac{i}{q^2-m_X^2}\langle0\vert J^X_\mu\vert  
X(q)\rangle\langle X(q)\vert J^{(cc)}_\nu\vert0\rangle\,,\nn\\ 
&=&\frac{i\lambda_X\lambda_W}{Q^2+m_X^2}\left(g_{\mu\nu}-\frac{q_\mu q_\nu} 
{m_X^2}\right) 
\end{eqnarray} 
where $q^2=-Q^2$,  and we  have used the  definition (\ref{2pmatrix}) 
and 
\begin{equation} 
  \langle0\vert J^X_\mu\vert X(q)\rangle=\lambda_X\epsilon_\mu(q)\,. 
\end{equation} 
The parameter defining the  coupling between the current $J^X_\mu$ and 
the $X$ meson has been  calculated in Ref.~\cite{x3872mix}, and its value 
is $\lambda_X = (3.6 \pm 0.9)\times10^{-3} \GeV^5$. 

In the QCDSR approach a Borel  transform to $Q^2\to M^2$ ($Q^2=-q^2$) is 
performed to  improve the matching between both  sides of the sum  rules. 
The Borel transform exponentially suppresses the contribution from excited  
states in the phenomenological side of the sum rule. In the OPE side the Borel 
transform suppresses the contribution from higher dimension condensates 
\cite{nnl}. After performing the Borel transform we  get in the 
structure $g_{\mu\nu}$: 
 
\begin{eqnarray}\lb{2psumrule} 
  \lambda_W\lambda_Xe^{-\frac{m_X^2}{M^2}}&=&-(\cos\alpha+\sin\alpha) 
\biggl(\sin\theta\,\Pi^{24}(M^2)\nn\\&+&\frac{\qq}{6\sqrt{2}}\cos\theta\, 
\Pi^{2,2}(M^2)\biggr)\,. 
\end{eqnarray} 
This  expression  is  analysed  numerically  to  obtain  the  coupling 
parameter  $\lambda_W$.  We  perform  the calculation  using the  same 
values for  the masses and QCD condensates  listed in \cite{x3872mix}, 
and in  the same  region in threshold  parameter $s_0$ and  Borel mass 
$M^2$  that we  have  used in  the  mass and  $\lambda_X$ analysis  in 
Ref.~\cite{x3872mix}, $\sqrt{s_0}=4.4\GeV$, $2.6  \GeV^2 \leq M^2 \leq 
3.0 \GeV^2$.  The mixing angles determined in the same reference are: 
\begin{equation}\lb{mixangles}5^\circ  \leq \theta \leq13^\circ\,, 
\quad\alpha=20^\circ\,.\end{equation} 
Taking into account the variation  in the Borel mass parameter and the 
mixing angle $\theta$, the result for the $\lambda_W$ parameter is: 
\begin{equation}\lb{lambdaW} 
  \lambda_W=(1.29\pm0.51)\GeV^2\,. 
\end{equation} 
\section{Three-point correlator} 
 
The  form factor of  the $B\to  K$ transition  matrix (\ref{3pmatrix}) can  
be evaluated from the three point correlator: 
\begin{equation} 
\Pi_{\mu}(p,p^\prime)=\!\!\int d^4x \,d^4y \,e^{i(p^\prime\cdot x-\,p 
\cdot y)}\langle0\vert T\{J_\mu^W(0)J_K(x)J^\dagger_B(y)\}\vert0\rangle 
\end{equation}

In  the   OPE  side  we  use   the  weak  current   $J^W_\mu$  defined  in 
(\ref{wcurrents})  and  the interpolating  currents  of  the  $B$ and  $K$ 
pseudoscalar mesons: 
\begin{eqnarray} 
  J_K(x)=i\,\bar{u}_a(x)\gamma_5 s_a(x)\,,\quad J_B=i\,\bar{u}_a(x) 
\gamma_5 b_a(x)\,. 
\end{eqnarray} 
We work at leading order  in $\alpha_s$ and consider condensates up to 
dimension 5 and terms linear in the mass of the $s$ quark. 
 
The phenomenological  side of the  sum rule is computed by inserting the 
intermediate states of the $B$ and $K$ mesons in the correlator: 
\begin{eqnarray} 
  \Pi^{\mathrm{phen}}_\mu=-\frac{f_Bf_Km_K^2m_B^2 }{m_b(m_s+m_u)}\frac{(f_+(t)(p+ 
p^\prime)_\mu+f_-(t)q_\mu)}{(p^2-m_B^2)(p^{\prime2}-m_K^2)}\,,\nn\\ 
\end{eqnarray} 
with $t=q^2=(p-p^\prime)^2$, using (\ref{3pmatrix}) and the following  
definitions: 
\begin{eqnarray} 
  \langle0\vert J_K\vert K(p^{\prime})\rangle&=&f_K\frac{m_K^2}{m_s+m_u}\nn\\ 
  \langle0\vert J_B\vert B(p)\rangle&=&f_B\frac{m_B^2}{m_b}\,. 
\end{eqnarray} 
Performing a double Borel  transform, $P^2\to M^2$ and $P^{\prime2}\to 
M^{\prime2}$, and matching  both sides of the sum rule,  we get in the 
structure     $(p_\mu+p^\prime_\mu)$      (with     $P^2=-p^2$,  
$P^{\prime2}=-p^{\prime2}$, $Q^2=-q^2=-t)$: 
\begin{widetext} 
\begin{eqnarray}\lb{sr3p} 
&&-\frac{f_Bf_Km_K^2m_B^2 f_+(t)}{m_b(m_u+m_u)}e^{-\frac{m_B^2}{M^2}- 
\frac{m_K^2}{M^{\prime2}}}=\frac{-1}{4\pi^2}\int_{s_{\min}}^{s_0} 
{\mathrm{ds}}\int_0^{u_0}{\mathrm{du}}\,\rho^{pert}(s,t,u)e^{-\frac{s}{M^2} 
-\frac{u}{M^{\prime2}}}+\frac{1}{2}\uu(m_b+m_s)e^{-\frac{m_b^2}{M^2}}\nn\\ 
&&-\frac{m_0^2\uu e^{-\frac{m_b^2}{M^2}}}{8M^4M^{\prime2}} \biggl((m_b^3 
+m_b^2m_s)(M^2+M^{\prime2})-M^2(m_b(M^2+2M^{\prime2}+t)+m_s(M^{\prime2} 
+2M^2+t))\biggr)\,, 
\end{eqnarray} 
where the perturbative contribution is given by 
\end{widetext} 
\begin{eqnarray} 
  && \rho^{\mathrm{pert}}(s,t,u)=\frac{3}{4{\lambda}^{\frac{3}{2}}(s,t,u)} 
\biggl[u(2m_b^2-s-t+u)\nn\\&\times&(2m_bm_s-s-u+t)+((s-m_b^2)(s-t+u)-2su) 
\nn\\&\times&(2m_bm_s-s+t-u)+(s+u-mb^2)\lambda(s,t,u)\biggr]\,.\nn\\ 
\end{eqnarray} 
 
The integration limit $s_{\min}$ is given by 
\begin{equation} 
  s_{\min}=m_b^2+\frac{m_b^2u}{m_b^2-t}\,. 
\end{equation} 
$s_0=(m_B+\Delta_s)^2$  and $u_0=(m_K+\Delta_u)^2$  are  the continuum 
threshold  parameters for the  $B$ and  $K$ respectively.   Note that, 
after  the double Borel  transform, the  contributions from  the quark 
condensate and mixed condensates of the  $s$ quark are eliminated. 
 
We use the following relation between the Borel masses $M^2$ and  
$M^{\prime2}$ \cite{gbbk}: 
 \begin{equation} 
   M^{\prime2}=\frac{0.64\GeV^2}{m_B^2-m_b^2}M^2\,.\end{equation} 
 
\subsection{Result for the form factor} 
 
The sum rules are analysed  numerically using the following values for  
quark masses  and QCD  condensates, and  for meson  
masses and decay constants  \cite{x3872,narpdg,pdg}:  
\beqa\label{qcdparam}  
&m_b(m_b)=4.7\,\GeV,\quad m_s=0.140 \GeV\nnb\\ 
&\qq=-(0.23\pm0.03)^3\,\GeV^3,\nnb\\  
&m_0^2=0.8\,\GeV^2,\nnb\\  
&m_{B} = 5.279 \GeV\,\quad m_{K} = 493.677 \MeV\nnb\\  
&f_{B} = 0.170 \GeV\,,\quad f_{K} = 0.160 \GeV  \,. 
\enqa  
We use  the value  of the  mixing angles $\al$  and $\theta$  given in 
(\ref{mixangles}).   For the  continuum threshold  parameters  we take 
$\Delta_s=\Delta_u=0.5\GeV$. 
 
\begin{figure}[t]  
\centerline{ 
\epsfig{width=0.5\textwidth,figure=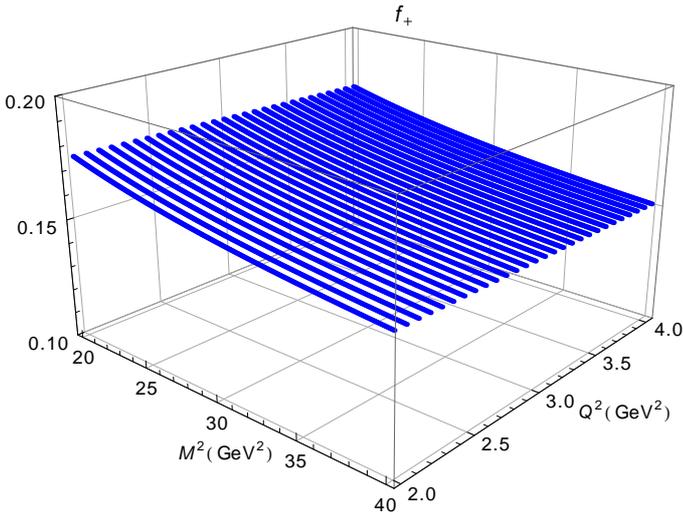} 
} 
\caption{\label{Q2vsM2}   The  plot   of  the   form  factor $f_+$ 
  calculated as a function of $Q^2$ and $M^2$ from Eq.~(\ref{sr3p}).} 
\end{figure}

In  Fig. \ref{Q2vsM2}  we  show the  plot  for the  form factor  $f_+$ 
calculated in the sum rules from Eq.~(\ref{sr3p}) as a function of the 
Borel  mass $M^2$  and the  momentum transfer  $Q^2$.  For  the region 
$M^2\ge20\GeV^2$  shown in  the  plot  the sum  rule  presents a  good 
stability. 
 
\begin{figure}[t] 
\centerline{ 
\hspace{-0.4cm} 
\epsfig{width=0.5\textwidth,figure=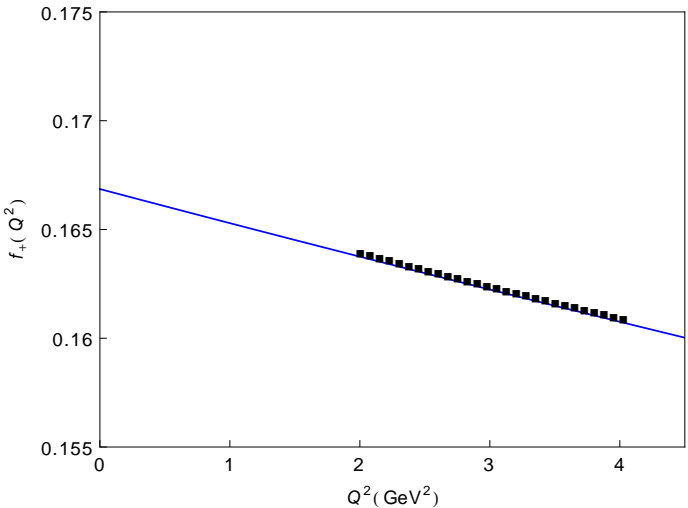} 
} 
\caption{\label{fit.form} Momentum dependence of  $C(Q^2)$ for 
$\Delta_s=\Delta_u=0.5 \GeV$. The solid line gives the  
parametrization of the QCDSR results (dots) through (\ref{fplus}).} 
\end{figure}

In the  following analysis, to  determine the $Q^2$ dependence  of the 
form factor, we  choose the Borel mass within  the region of stability 
around  the  mass of  the  $B$  meson,  $26 \GeV^2\leq  M^2\leq30\GeV^2$.  
We  fit 
(\ref{sr3p}) within  the stable region  by matching both sides  of the 
sum  rule.  In  Fig.  \ref{fit.form}  we show,  through the  dots, the 
QCDSR results for  the form factor $f_+(Q^2)$ as  a function of $Q^2$. 
The   numerical  results   can   be  well   fitted   by  a   monopolar 
parametrization (shown by the solid line in Fig. (\ref{fit.form})): 
\begin{equation}\lb{fplus} 
  f_+(Q^2)=\frac{(17.55\pm0.04) \GeV^2}{(105.0\pm1.76)\GeV^2+Q^2}\,. 
\end{equation} 

For the decay width calculation, we  need the value of the form factor 
at  $Q^2=-m_X^2$,  where  $m_X$  is  the mass  of  the  off-shell  $X$ 
meson. Therefore we have: 
\begin{equation}\lb{fpluspolo} 
  f_+(Q^2)\vert_{Q^2=-m_X^2}=0.195\pm0.003\,. 
\end{equation} 

\subsection{The decay width $B\to XK$} 
 
To  determine  the  the  decay  width we  insert,  in  the  expression 
(\ref{eqwidth}), the parameter $\lambda_W$ (\ref{lambdaW}) obtained in 
the two-point  sum rule calculation, and  the value of  the form factor 
$f_+$,  from  the  three-point  sum   rule,  at  the  $X$  meson  pole 
(\ref{fpluspolo}).   The  branching   ratio  is  therefore  calculated 
dividing   the  result   by  the   total  width   of  the   $B$  meson 
$\Gamma_{\mathrm{tot}}=\hbar/\tau_B$: 
\begin{equation}\lb{result} 
  \mathcal{B}(B\to X(3872)K)=(1.00\pm0.68)\times10^{-5}\,, 
\end{equation} 
where   we   have   used   the    mean   life   of   the   $B$   meson 
$\tau_B=1.638\times10^{-12}$  s,  the  CKM parameters  $V_{cs}=1.023$, 
$V_{cb}=40.6\times10^{-3}$  \cite{pdg}, and  the  Wilson coefficients 
$C_1(\mu)=1.082$,   $C_2(\mu)=-0.185$,  computed   at   $\mu=m_b$  and 
$\bar{\Lambda}_{\mathrm{MS}}=225\MeV$    \cite{buras}.    The   result 
(\ref{result})  is  in agreement  with  the  experimental upper  limit 
(\ref{branching}). 
 
For completeness  we also  compute the branching  ratio for the  $X$ as 
pure $\bar{c}c$ and  molecular states.  We choose the  mixing angle as 
$\theta=90^\circ$ and  $\theta=0^\circ$ in Eq.   (\ref{field}), and we 
get respectively for the pure molecule and pure charmonium: 
\begin{equation}\lb{resmol} 
  {\mathcal{B}}(B\to X_{\mathrm{mol}}K)=(0.38\pm0.06)\times10^{-6}\,, 
\end{equation} 
\begin{equation}\lb{resccbar} 
  {\mathcal{B}}(B\to X_{\bar{c}c}K)=(2.68\pm0.50)\times10^{-5}\,. 
\end{equation} 
Comparing the results  for the pure states with the  one for the mixed 
state (\ref{result}), we can see that the branching ratio for the pure 
molecule is  one order smaller,  while the pure charmonium  is larger. 
 
The result for the pure molecular state in Eq.~(\ref{resmol}) can be compared 
with the one from Ref.~\cite{bxk1}. In this work the authors study the 
production of  the $X$  as a  molecular state in  the decay  $B\to XK$ 
through  the coalescence  of  charm mesons.   The  calculation of  the 
branching ratio is strongly dependent on the choice of parameters, and 
it   is  found   to  be   of   order  $10^{-4}$   to  $10^{-6}$.    In 
Ref.~\cite{bxk3}, the  production of the $X$ as  a $2^3P_1$ charmonium 
state is studied in pQCD, and  their result for the branching ratio is 
$7.88^{+4.87}_{-3.76}\times10^{-4}$, which is  an order bigger than our 
result for the pure charmonium (\ref{resccbar}). 
 
\section{Summary and conclusions} 
 
We have presented a QCDSR  analysis of the production of the $X(3872)$ 
state considering  a mixed  charmonium-molecular current in  the decay 
$B\to XK$.  We find that the sum rules result in Eq.~(\ref{result}), 
obtained by using the factorization hypothesis, is 
smaller, but  compatible with the experimental  upper limit.  Since, 
it is known that  non-factorizable contributions may play an important 
role  in   hadronic  decays  of   $B$  mesons ~\cite{Khodjamirian},  
our result can be interpreted as a  lower limit for the branching ratio. 
 
This  result  was  obtained   by  considering  the  mixing  angles  in 
Eqs.~(\ref{4mix}) and (\ref{field})  with the values $\al=20^\circ$ and 
$5^\circ  \leq \theta  \leq  13^\circ$.  The  present  result is  also 
compatible with previous analysis of the mass of the $X$ state and the 
decays    into    $J/\psi\pi^0\pi^+\pi^-$    and    $J/\psi\pi^+\pi^-$ 
\cite{x3872mix},    and   the    radiative   mode    $\gamma   J/\psi$ 
\cite{x3872rad}, since  the values of  the mixing angles used  in both 
calculations are the  same.  It is important to  mention that there is 
no  new free  parameter in  the present  analysis and,  therefore, the 
result   presented  here   strengthens  the   conclusion   reached  in 
Refs.~\cite{x3872mix,x3872rad} that the $X(3872)$ is probably a mixture 
between           a           $c\bar{c}$           state           and 
$D^0\bar{D}^{*0},~\bar{D}^0{D}^{*0},~D^+D^{*-}$     and    $D^-D^{*+}$ 
molecular states. From Eq.~(\ref{field}) one may be tempted to say that 
the $c\bar{c}$ component of the state described by such current 
is dominant  ($\sim$97\%), like  
the conclusion presented in Ref.~\cite{x3872mix}. However, from a closer 
look at Eq.~(\ref{curr2}), one can see that the $c\bar{c}$ component 
of our current is already multiplied by a dimensional parameter, the 
quark condensate, in order to have the same dimension of the 
molecular part of the current. Therefore, it is not clear that only the 
angle in Eq.~(\ref{field}) determines the percentage of each component.  
To try to evaluate the importance of each part of the current it is 
better to analyse the results obtained with each component, like the 
results presented in Eqs.~(\ref{resmol}) and (\ref{resccbar}). From these  
results we see that the $c\bar{c}$ part of the state plays a very important 
role in the determination of the branching ratio. By the other hand, in the 
decay $X\to J/\psi\pi^+\pi^-$ and $X\to J/\psi\pi^+\pi^-\pi^0$, the width  
obtained in our approach for a pure $c\bar{c}$ state is \cite{x3872mix}: 
\begin{equation}\lb{xppcc} 
  \Gamma(X_{\bar{c}c}\to J/\psi n\pi)=0\,, 
\end{equation} 
and, therefore, the molecular part of the state is the only one  
that contributes to this decay, playing an essential role  
in the determination of this decay width. Also, for a pure $c\bar{c}$ state  
one gets: 
\begin{equation} 
M_{X_{c\bar{c}}}=(3.52\pm0.05)~\mbox{GeV}, 
\end{equation} 
from where one sees again that the molecular part of the state plays a 
very important role in the determination of its mass. 
 
Therefore, although we cannot determine the percentages of the $c\bar{c}$ 
and the molecular components in  the $X(3872)$, we may say that both 
components are extremely important, and that, in our approach, it is not  
possible to explain all the experimental data about the $X(3872)$ with 
only one component. 
 
\section*{Acknowledgment} 
 
\noindent  
The authors would like to thank M.E. Bracco for discussions. 
 This work has been partially supported by FAPESP and CNPq.

\end{document}